\documentclass{article}
\usepackage{cite}
 \usepackage{graphicx} \usepackage{amsmath}

\suppressfloats[t]

\begin{document} \large

\begin{center}
{\bf Wave Turbulence of a Liquid Surface in an External Tangential Electric Field}\\[1.0ex]

\vspace{2mm}
{\it Evgeny A. Kochurin$^1$ }\\[1.0ex]
{$^1$Institute of Electrophysics, Ural Division,
Russian Academy of Sciences, 106 Amundsen Street, 620016 Ekaterinburg, Russia\\
}
\end{center}

\vspace{2mm}
\begin{abstract}

\large
A direct numerical simulation of the interaction of plane capillary waves on the surface of a liquid dielectric in an external tangential electric field taking into account viscous forces has been performed. It has been shown that the interaction of counter-propagating nonlinear waves can generate a direct energy cascade. In
the quasistationary energy dissipation regime, probability density functions for angles of inclination of the boundary tend to a Gaussian distribution and the shape of the boundary becomes complex and chaotic. The spectrum of the surface perturbations in this regime is described by a power law $k^{-5/2}$. The energy spectrum
has the form $k^{-3/2}$, which coincides with the Iroshnikov-Kraichnan energy spectrum and indicates that the observed wave turbulence of the liquid surface and the weak magnetohydrodynamic turbulence of interacting Alfv\'en waves have a related nature.

\emph{The original article is published in JETP Letters, 2019, Vol. 109, No. 5, p. 303.}
\end{abstract}

\vspace{2mm}

It is known that the dynamics of nonlinear capillary waves can be fairly complex. Zakharov and Filonenko \cite{Zakh} showed that the weak turbulence of capillary
waves can be developed at the boundary of a liquid. Within a weakly nonlinear model, under the assumption that the amplitudes of capillary waves are
small, they demonstrated that the spectrum of surface perturbations $I_k$ in a small-scale region (the gravitational force is negligibly weak) has the power-law form
\begin{equation}\label{ZF}I_k=<|\eta_k|^2>\sim C k^{-(11/4+d)},\end{equation}
where $\eta_k$ is the Fourier transform of the function $\eta$ determining the shape of the boundary of the liquid, $k$ is the length of the wave vector, $C$ is a constant, $d$ is the number of spatial variables of the function $\eta$, and the angle brackets mean angular averaging.

The first numerical experiments definitely confirming the existence of spectrum (1) were reported in \cite{Push}, and this result was confirmed by physical experiments in  \cite{mezhov1,mezhov2}. Simulating the complete strongly nonlinear hydrodynamic equations, the authors \cite{falcon1}  obtained the Zakharov-Filonenko turbulent spectrum. Waves with a large wavelength, whose evolution is determined primarily by the gravitational force, can exhibit the Phillips \cite{phil} and Kolmogorov-Zakharov  spectra \cite{KZ}. Thus, the effect of capillary and gravitational forces on the development of turbulence of surface waves is well studied (see review [8]).

The situation changes significantly in the presence of external electromagnetic fields, which can strongly affect the evolution of the boundary \cite{Me1}. The dynamics of coherent structures on the liquid surface in an electric field has been studied in detail (see, e.g., \cite{zhakin, koulova,tao1,tao2,zu01,zu02, gao, wang}). The chaotic dynamics of the surface of waves in the external field almost has not been studied (except for two experimental works \cite{falcon2,falcon3}). For this reason, the study of the wave turbulence of the liquid surface appearing in a high external electric field is very important. This work is devoted to such a study.

A numerical simulation of weakly nonlinear dynamics of plane-symmetric capillary waves on the surface of a dielectric liquid with a high dielectric constant
in an external tangential electric field is presented in this work. In a high electric field, when the effect of capillary and gravitational forces is negligibly small, the exact analytical solutions of this problem were first obtained by Zubarev \cite{zu1}. These solutions describe the dispersionless propagation of nonlinear
surface waves with an arbitrary shape in the direction of the field or against it \cite{zu2,zu3}. The interaction of counterpropagating waves results in the appearance of singular points at the boundary, at which the field strength and the velocity of the liquid are discontinuous and the curvature of the surface increases infinitely \cite{ko1,ko2,ko3}. The calculations performed in this work indicate that the previously observed singularities are smoothed when the surface tension and viscous forces are taken into account. Furthermore, a direct energy cascade and the chaotization of evolution of the system can be observed in the regime of propagation of electrocapillary waves. In the quasistationary energy dissipation regime, the spectrum of surface perturbations tends to a power law, $k^{-5/2}$.

We consider a potential flow of an ideal incompressible dielectric liquid with infinite depth and a free surface in a uniform horizontal external electric field.
Since the problem under consideration is anisotropic because of the existence of the separated direction of the electric field, we consider only plane symmetric
waves propagating in the direction parallel to the external field. Let the field strength be directed along the $x$ axis (correspondingly, the $y$ axis of the Cartesian coordinate system is perpendicular to it) and have the magnitude $E$. The shape of the boundary in an unperturbed state corresponds to $y=0$.

The dispersion relation for linear waves at the boundary of the liquid has the form [9]
\begin{equation}\label{disp}\omega^2=gk+\frac{\varepsilon_0 \varepsilon E^2}{\rho}k^2+\frac{\sigma}{\rho}k^3,\end{equation}
where $\omega$ is the frequency; $g$ is the gravitational acceleration; $\varepsilon_0$ is the electric constant; $\varepsilon$ ($\varepsilon\gg1$) and $\rho$ and ñ are the dielectric constant and density of the liquid, respectively; and $\sigma$ is the surface tension coefficient. For further analysis, it is convenient to introduce the dimensionless variables
$$y\to y/k_0,\quad x\to x/k_0,\quad t\to t \cdot t_0,\quad E\to\beta \cdot E_0,$$
where $k_0=(g \rho/ \sigma)^{1/2}$, $t_0=(\sigma/g^3 \rho)^{1/4}$, and $E_0^2= (\sigma g \rho)^{1/2}/\varepsilon_0\varepsilon$ are the characteristic values of the wavenumber,
time, and electric field strength, respectively, and $\beta=E/E_0$ is the dimensionless electric field strength. In particular, for water
($\varepsilon\approx 81$), $\lambda=2\pi/k_0\approx 1.7$ cm, $t_0 \approx 0.01$ s, and $E_0\approx$ 1.9 kV/cm.

Below, we consider the region of electrocapillary waves $\beta^2+k\gg 1/k$, i.e., wavelengths for which the effect of the gravitational force can be neglected. The dispersion relation (2) can be represented in the dimensionless form
\begin{equation}\label{disp2}\omega^2=(\beta^2+k)k^2.\end{equation}

The equations of motion of the boundary of the liquid up to the second order terms can be represented in the form \cite{zu1}
$$\psi_t=\eta_{xx}+\frac{1}{2}\left[\beta^2(\hat k \eta^2-\eta_x^2) +\hat k \psi^2-\psi_x^2\right]+$$
\begin{equation}\label{eq1}+\beta^2 \left[-\hat k \eta +\hat k(\eta\hat k \eta)+\partial_x(\eta \eta_x)\right]+\hat D_k \psi,\end{equation}

\begin{equation}\label{eq2}\eta_t=\hat k \psi- \hat k(\eta \hat k \psi)-\partial_x(\eta \psi_x)+\hat D_k \eta,\end{equation}
where $\psi$ is the function determining the potential of the velocity of the liquid at the boundary, $\hat k$ is the integral
operator having the form: $\hat k f_k=|k| f_k$ in the Fourier representation, and the operator $\hat D_k$ describes viscosity
and is defined in the $k$ space as
$$\hat D_k=-\gamma (|k|-|k_d|)^2, \quad |k|\geq |k_d|;\quad  \hat  D_k=0,\quad |k|< |k_d|.$$
Here, $\gamma$ is a constant, and $k_d$ is the wavenumber determining the spatial scale at which the energy dissipation occurs. This definition of viscosity is standard at the simulation of the wave turbulence of the liquid surface in the absence of an electric field \cite{korot}.

Equations (4) and (5) are Hamiltonian and can be derived as the variational derivatives
$$\frac{\partial \psi}{\partial t}=-\frac{\delta H}{\delta \eta},\qquad \frac{\partial \eta}{\partial t}=\frac{\delta H}{\delta \psi}.$$
Here,
$$H=H_0+H_1=\frac{1}{2}\int \left(\psi \hat k \psi+\beta^2 \eta \hat k \eta+\eta_x^2 \right)dx-$$
$$-\frac{1}{2}\int \eta\left([\hat k \psi^2-\psi_x^2]+\beta^2[\hat k \eta^2-\eta_x^2]\right)dx,$$
is the Hamiltonian of the system specifying the total energy. Here, $H_0$ and $H_1$ are the linear and quadratic terms, respectively. For the weakly nonlinear model, the condition $H_1/H_0\ll 1 $ should be satisfied.

For high fields ($\beta\gg1$), i.e., in the absence of capillary and viscous forces, equations of motion of the boundary (4) and (5) allow a pair of exact solutions in the form
$$\psi=\pm \beta \hat H \eta,$$
where $\hat H $is the Hilbert transform operator: $\hat H f_k=i \mbox{sign}(k) f_k$. These solutions are obtained by Zubarev \cite{zu1}. As mentioned above, they describe the propagation of waves without distortions in the direction of the field or against it (depending on the sign in the above formula).

To minimize the effect of coherent structures (collapses or solitons), initial conditions for Eqs. (4) and  (5) are taken in the form of two counter-propagating
interacting wavepackets:
\begin{equation}\label{IC}\eta_1(x)=\sum \limits_{i=1}^{3}a_i\cos(k_i x),\quad\eta_2(x)=\sum \limits_{i=1}^{3}b_i\cos(p_i x),
\end{equation}
$$\eta(x,0)=\eta_1+\eta_2,\qquad \psi(x,0)=\beta(\hat H \eta_1-\hat H \eta_2).$$

The numerical experiments were performed for three field strengths $\beta^2_1=35$, $\beta^2_2=30$, $\beta^2_3=25$. The spatial derivatives and integral operators were calculated using pseudo-spectral methods with the total number $N$ of harmonics, and the time integration was performed by the fourth-order explicit Runge–Kutta
method with the step $dt$. To stabilize the numerical scheme, the amplitudes of higher harmonics with a wavenumber above $k_s$ were equated to zero at each step in time; i.e., a low-frequency filter was used (similar to \cite{Push}).   

A qualitatively similar behavior was observed in all three experiments. The calculations did not involve the mechanical pumping of the energy of the system. Hence, the average steepness of the boundary $\mu= <|\eta_x|>$  was determined only by initial conditions (6). The calculation was performed with the parameters $N=2048$, $dt=10^{-5}$, $\gamma=10$, $k_d=500$, $k_s=624$. All numerical experiments demonstrate the formation of an inertial interval in the wavenumber range $30< k< 150$. A decrease in the field strength is accompanied by an increase in the nonlinearity threshold necessary for the formation of a direct energy cascade. In particular, three experiments gave the average steepness of the boundary $\mu_1\approx0.116$, $\mu_2\approx0.121$, $\mu_3\approx0.127$, respectively.

We present in more detail the results of calculation for $\beta^2_1=35$ (at high field strengths, experiments demonstrated strong singularities, which can correspond to the formation of vertical liquid jets \cite{ko4}). The amplitudes and wavenumbers in initial conditions (6) were chosen empirically to minimize the deviation of
the model from weakly nonlinear. The calculations reported below were performed with the parameters
$$a_1=0.01125, a_2=0.0165, a_3=0.01,$$
$$b_1=0.0125, b_2=0.0145, b_3=0.01,$$
$$ k_1=4, k_2=6, k_3=8, \quad p_1=5, p_2=7, p_3=9.$$

Figure 1a shows the time dependence of the relative change $\Delta H$ in the energy of the system. It is seen that, beginning with a certain time, energy dissipation
occurs with almost a constant rate. This behavior correlates well with the time dependence of the space-period-averaged magnitude $<|K|>$  of the curvature of
the boundary shown in Fig. 1b. The magnitude $<|K|>$ reaches a maximum at $t\approx 3.2 \cdot 10^3 $ and, then, decreases almost monotonically. Figure 1c shows the
time dependence of the ratio $H_1/H_0$, characterizing the nonlinearity level of the system. The plot does not demonstrate such a transition regime, and the ratio $H_1/H_0$ remains small during the entire integration time interval. The average nonlinearity level in the process of evolution of the system was small, about $1.5\cdot10^{-2}$.

\begin{figure}[h]
\center{\includegraphics[width=.5\linewidth]{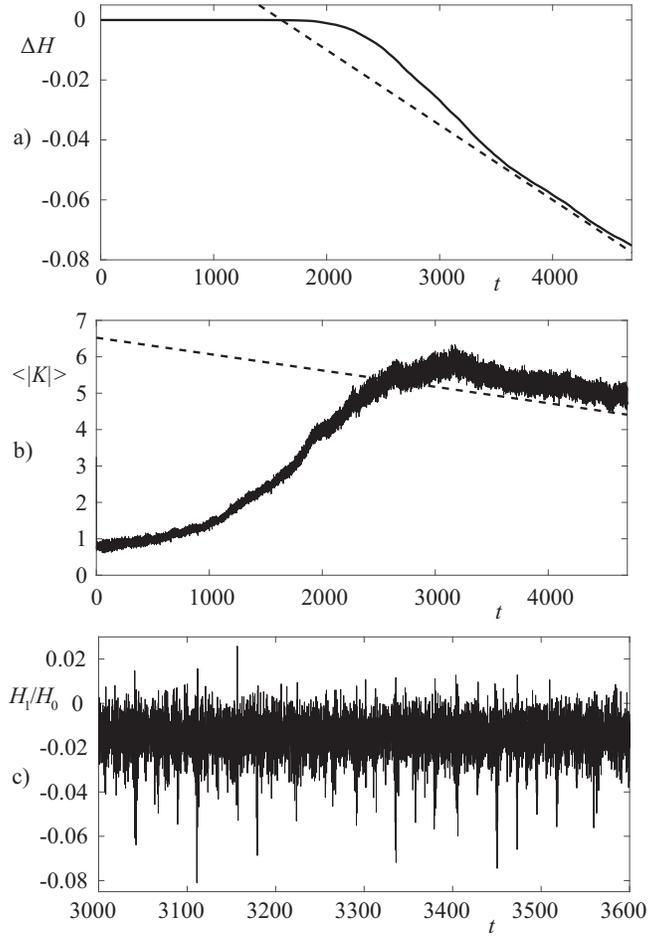}}
\caption{Time dependences of (a) relative change in the energy of the system, (b) average magnitude of the curvature of the boundary, and (c) nonlinearity level.}
\label{buble}
\end{figure}

Figure 2 shows the shape of the boundary of the liquid at the initial time and at $t=3.16\cdot 10^3$. It is seen that this boundary in the quasistationary energy dissipation regime has a complex irregular shape. In this case, the probability density functions for angles of inclination of the boundary become very close to a
Gaussian distribution (see Fig. 3). This behavior indicates the absence of strong space–time correlations and, consequently, the possible appearance of the
Kolmogorov spectrum of wave turbulence.

\begin{figure}[h]
\center{\includegraphics[width=0.6\linewidth]{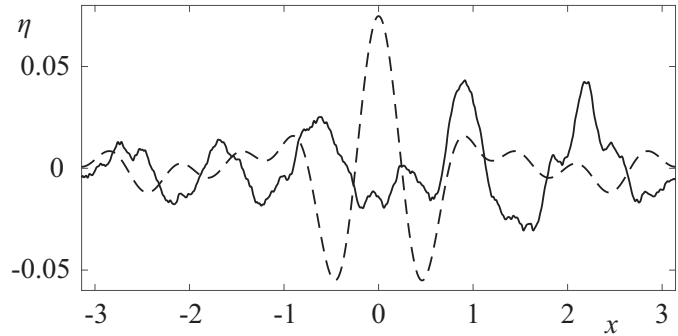}}
\caption{Shape of the boundary of the liquid at (dashed line) the initial time and (solid line) $t=3.16\cdot 10^3$.}
\label{buble}
\end{figure}

\begin{figure}[h]
\center{\includegraphics[width=0.6\linewidth]{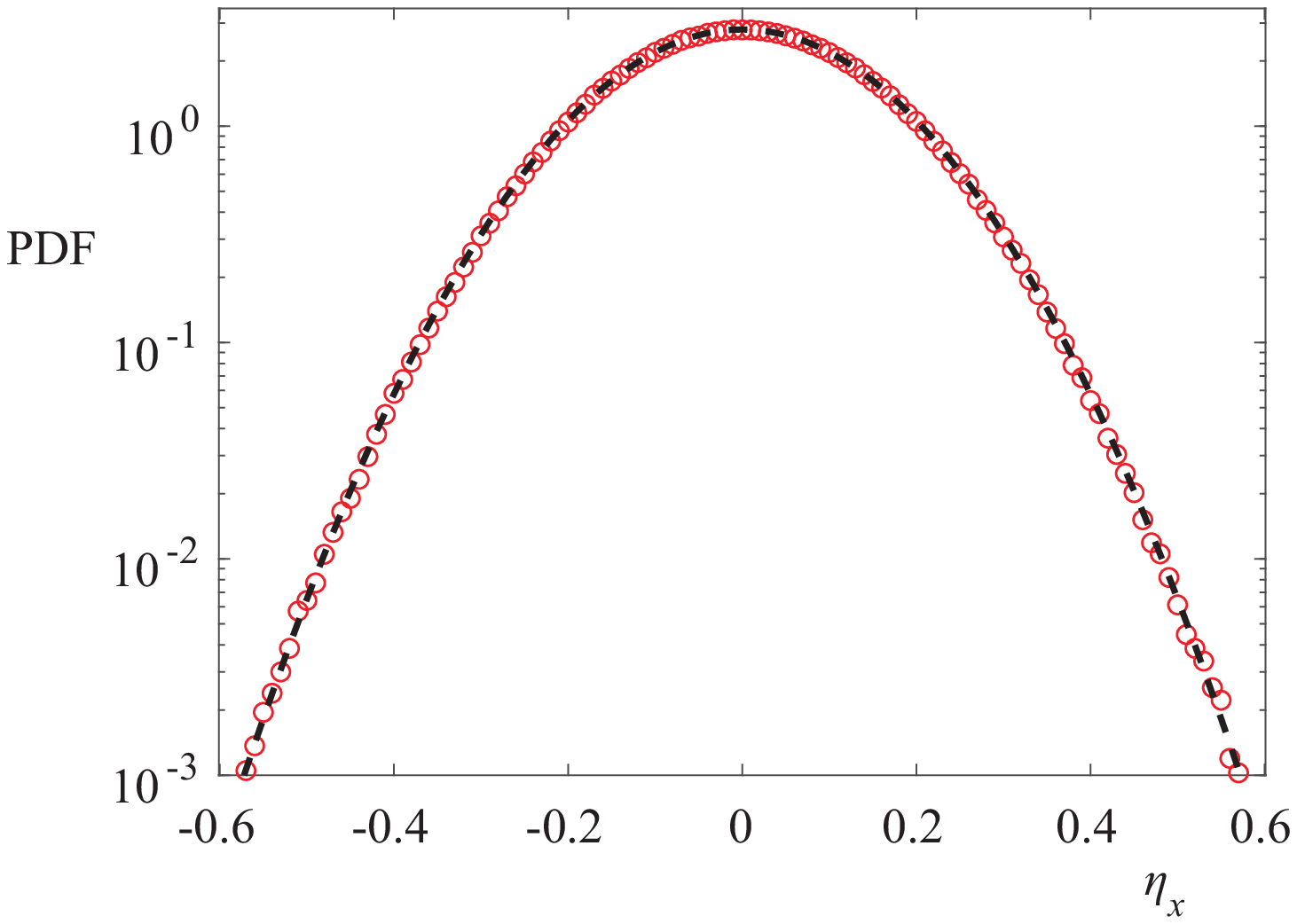}}
\caption{(Color online) Probability density function for the steepness of the boundary. The dashed line is the Gaussian distribution.}
\label{buble}
\end{figure}

Figure 4 shows the time-averaged spectrum of surface perturbations of the boundary of the liquid:
$$I_{\beta}=(T)^{-1}\int\limits_{t_1}^{t_2}|\eta_k|^2dt,$$
where $T=t_2-t_1$ is the time averaging period (averaging was performed in a stationary regime).

\begin{figure}[h]
\center{\includegraphics[width=0.6\linewidth]{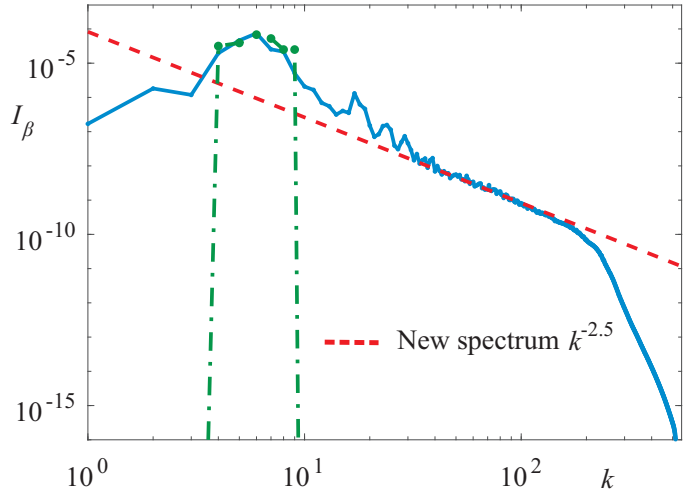}}
\caption{(Color online) Spectra of (blue solid line) surface perturbations $I_{\beta}$, (green dash-dotted line) initial perturbation, and (red dashed line) $k^{-2.5}$.}
\label{buble}
\end{figure}

As seen in Fig. 4, the resulting spectrum of surface perturbations is far from the classical Zakharov-Filonenko spectrum and can be approximated with a high accuracy by a power law $k^{-2.5}$. It is remarkable that the slope of the spectrum in the wave turbulence regime is the same for all performed calculations. Figure
5 shows the compensated spectra for three numerical experiments with different â values. It is seen that the empirical dependence obtained describes the
numerical experiments very well. The results of calculations can generally be interpreted as the detection of a new spectrum of wave turbulence of the liquid surface
different from classical spectra for capillary and gravitational waves.

\begin{figure}[h]
\center{\includegraphics[width=0.6\linewidth]{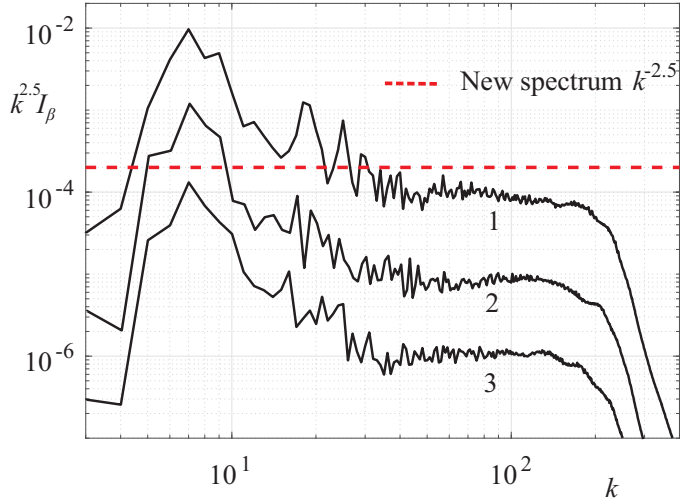}}
\caption{Ðèñ. 5. (Color online) Compensated spectra of surface perturbations
for various electric field strengths: ($1$) $k^{2.5}I_{\beta}$ ($\beta^2=35$), ($2$) $0.1 k^{2.5}I_{\beta}$ ($\beta^2=30$), ($3$) $0.01k^{2.5}I_{\beta}$  ($\beta^2=25$), and (red dashed straight line) $k^{-2.5}$.}
\label{buble}
\end{figure}

The reported calculations were performed for quite high external electric fields. A decrease in the field strength is accompanied by an increase in the nonlinearity
level necessary for the implementation of the direct cascade energy. The Zakharov-Filonenko spectrum of turbulence (1) was not obtained in the case of zero external electric field. This can be due to the absence of conditions for three-wave resonance of capillary waves in one-dimensional geometry. Indeed, it is easy to show that the conditions for three-wave resonance of capillary waves ($\omega^2=k^3$),
\begin{equation}\label{reson}\omega=\omega_1+\omega_2,\qquad k=k_1+k_2,\end{equation}
cannot be satisfied in one-dimensional geometry. In the presence of a high electric field, i.e., at $\beta\gg1$, the
dispersion relation (3) for low wavenumbers $k<\beta^2$ can be written in the form
$$\omega^2=\beta^2 k^2.$$
For this relation, conditions (7) are satisfied at any $k$ values. Thus, the probability of satisfaction of the conditions for three-wave resonance of electrocapillary waves (3) increases with the field strength.

It is noteworthy that wave turbulence studied in this work is very similar to magnetohydrodynamic (MHD) wave turbulence of interacting Alfv\'en waves. It is known that waves in an ideally conducting liquid can propagate without distortions along the direction of the external magnetic field. Interaction is possible only between counterpropagating waves, and this interaction is elastic \cite{mhd0}. Surface waves in the high electric field regime studied in this work have the same properties \cite{ko1}. The classical result of the study of MHD wave turbulence is the Iroshnikov-Kraichnan spectrum \cite{mhd1,mhd2}. We briefly illustrate the derivation of this spectrum and show that the resulting empirical dependence  $I_{\beta}\sim k^{-5/2}$ can be closely related to MHD wave turbulence.

In the presence of a high field, the time of interaction $\tau$ between counter-propagating "quasiparticles" (harmonics) of the magnetic field (in our case, the
electric field) with the wavelength $\lambda\sim k^{-1}$ and velocity $v_{k}$ can be estimated as
$$\tau\sim 1/ k V_A,$$
where $V_A=\sqrt{\varepsilon_0 \varepsilon E^2/\rho}$ is an analog of the Alfv\'en velocity for our problem. Velocity perturbations are small, i.e., $v_k/V_A\ll 1$. The velocity $v_{k}$ is related to the spectral density of the kinetic energy $E_T(k)$ in the one-dimensional case as $v_{k}\sim [E_T(k)k]^{1/2}$. For the elastic interaction of waves, the spectral densities of the kinetic $E_{T}$ and potential (electrostatic) $E_{U}$ energies are proportional to each other $E_{T}\sim E_{U}$. At an elementary interaction event, the wavepacket is distorted $\Delta v_{k}\sim v_{k}^2/V_A$ (this expression follows from the assumption of the dominant effect of quadratic terms). The number of ``collisions'' $N_{int}$ necessary for a significant distortion of the wavepacket is estimated as \cite{mhd3}
$$N_{int}=\frac{V_A}{\Delta v_{k}}\sim \left(\frac{V_A}{v_{k}}\right)^2\gg1.$$

Then, the energy redistribution time can be introduced as $\tau_E\sim N_{int}/kV_A$. The energy dissipation rate $s$ should be independent of the wavenumber (condition of the local energy transfer)
$$v_{k}^2/\tau_E=s.$$
Substituting the expression for $\tau_E$ into this equality, one can obtain the spectral distributions for the velocity $v_{k}$ and, therefore, for the kinetic $E_{T}$ and potential $E_{U}$ energies in the turbulent regime,
\begin{equation}\label{IKspectr} v_{k}^2\sim(s V_A)^{1/2}k^{-1/2}, \quad E_{U}\sim E_{T}\sim (s V_A)^{1/2}k^{-3/2}.
\end{equation}
Expressions (\ref{IKspectr}) are classical Iroshnikov-Kraichnan spectra. In the case of a high field, the velocity fluctuations  $v_{k}$ are related to surface perturbations $\eta_k$ as $v_{k}\sim V_A k \eta_k$. Consequently, the desired spectrum of surface perturbations is obtained from Eq. (8) in the form \begin{equation}\label{spec}<|\eta_k|^2>=I_{\beta}\sim s^{1/2}V_A^{-3/2}k^{-5/2}.\end{equation}

Thus, the spectrum $I_{\beta}$  obtained in numerical experiments is in qualitative agreement with Eq. (9), which directly follows from the Iroshnikov-Kraichnan energy spectrum (8) for MHD wave turbulence. To conclude, the results obtained in this work are applicable for a ferromagnetic liquid with a high dielectric constant in an external magnetic field. These results, together with the data obtained in \cite{falcon2, falcon3}, indicate the existence of a new spectrum of the wave turbulence of the liquid surface in an external magnetic (electric) field

I am deeply grateful to N.M. Zubarev and N.B. Volkov for stimulating discussions. This work was supported in part by the Ural Branch of the Russian Academy of Sciences (project no. 18-2-2-15) and the Russian Foundation for Basic Research (project no. 17-08-00430).

\begin{flushright}{ Translated by R. Tyapaev}\end{flushright}


\begin{thebibliography}{99}

\bibitem{Zakh}
V. E. Zakharov and N. N. Filonenko, J. Appl. Mech. Tech. Phys. V. 4, p. 506, (1967).
    
\bibitem{Push}
A.N. Pushkarev, V.E. Zakharov, Phys. Rev. Lett. \textbf{76} (18), 3320 (1996).

\bibitem{mezhov1}
M. Yu. Brazhnikov, G. V. Kolmakov, A. A. Levchenko, and L. P. Mezhov-Deglin, JETP Lett. 73, 398 (2001).

\bibitem{mezhov2}
M.Yu. Brazhnikov, G.V. Kolmakov,  A.A. Levchenko,  L.P. Mezhov-Deglin, Europhys. Lett. \textbf{58} (4), 510 (2002).

\bibitem{falcon1}
L. Deike, D. Fuster, M. Berhanu, E. Falcon, Phys. Rev. Lett. \textbf{112}, 234501 (2014)

\bibitem{phil}
O.M. Phillips, J. Fluid Mech. \textbf{4}, 426 (1958).

\bibitem{KZ}
V.E. Zakharov, N.N. Filonenko, Sov. Phys. Dokl. \textbf{11}, 881 (1967).

\bibitem{naz}
S. Nazarenko, S. Lukaschuk, Annu. Rev. Condens. Matter Phys. \textbf{7}, 61 (2016).

\bibitem{Me1}
J.R. Melcher, Phys. Fluids \textbf{4}, 1348 (1961).

\bibitem{zhakin}
A. I. Zhakin, Phys. Usp. 56, 141 (2013).

\bibitem{koulova}
D. Koulova, H. Romat, C.L. Louste, IEEE Trans. Diel. Electr. Insul. \textbf{25} (5), 1716 (2018).

\bibitem{tao1}
B. Tao, D.L. Guo, Comput. Math. Appl. \textbf{67}, 627 (2014).

\bibitem{tao2}
B. Tao, Comput. Math. Appl. \textbf{76}, 799 (2018).

\bibitem{zu01}
N.M. Zubarev, Phys. Rev. E \textbf{65}, 055301 (2002).

\bibitem{zu02}
N.M. Zubarev, Phys. Fluids \textbf{18}, 028103 (2006).

\bibitem{gao}
T. Gao, P.A. Milewski, D.T. Papageorgiou, J.-M. Vanden-Broeck, J. Eng. Math. \textbf{108}, 107 (2018).

\bibitem{wang}
Z. Wang, Proc. R. Soc. A  \textbf{473}, 20160817 (2017).

\bibitem{falcon2}
S. Dorbolo, E. Falcon, Phys. Rev. E \textbf{83}, 046303 (2011).

\bibitem{falcon3}
F. Boyer, E. Falcon, Phys. Rev. Lett. \textbf{101}, 244502 (2008).

\bibitem{zu1}
N.M. Zubarev, Phys. Lett. A \textbf{333}, 284 (2004).

\bibitem{zu2}
N. M. Zubarev and O. V. Zubareva, Tech. Phys. Lett. 32, 886 (2006).

\bibitem{zu3}
N. M. Zubarev, JETP Lett. 89, 271 (2009).

\bibitem{ko1}
N. M. Zubarev and E. A. Kochurin, JETP Lett. 99, 627 (2014).

\bibitem{ko2}
E. A. Kochurin, J. Appl. Mech. Tech. Phys. 59, 79
(2018).

\bibitem{ko3}
E.A. Kochurin, N.M. Zubarev, IEEE Trans. Diel. Electr. Insul. \textbf{25} (5), 1723 (2018).

\bibitem{korot}
A.O. Korotkevich, A.I. Dyachenko, V.E. Zakharov, Physica D \textbf{321-322}, 51 (2016)

\bibitem{ko4}
E.A. Kochurin, N.M. Zubarev, J. Phys.: Conf. Ser.  \textbf{946}, 012021 (2018).

\bibitem{mhd0}
P. Goldreich and S. Sridhar, Astrophys. J. \textbf{438}, 763 (1995).

\bibitem{mhd1}
P. S. Iroshnikov, Sov. Astron. 7, 566 (1963).

\bibitem{mhd2}
R.H. Kraichnan, Phys. Fluids \textbf{8}, 1385 (1965).

\bibitem{mhd3}
P. Goldreich, S. Sridhar, Astrophys. J. \textbf{485}, 680 (1997).

\end{thebibliography}
\end{document}